\documentclass[prd,tightenlines,nofootinbib,superscriptaddress]{revtex4-2}

\usepackage{amsfonts,amsmath,amssymb,amsthm,bbm,hyperref}
\usepackage{enumitem}

\usepackage{mathptmx}       
\usepackage{helvet}         
\usepackage{courier}        
\usepackage{type1cm}        
\usepackage{graphicx}        
\usepackage{hyperref}        
\usepackage{tikz}
\usetikzlibrary{calc}
\usetikzlibrary{decorations.pathmorphing}
\usetikzlibrary{shapes.geometric}
\usetikzlibrary{arrows,decorations.markings}
                       
\newcommand{\C}{{\mathbb C}}
\newcommand{\N}{{\mathbb N}}

\newcommand{\cE}{{\mathcal E}}

\newcommand{\cH}{{\mathcal H}}

\newcommand{\cP}{{\mathcal P}}

\newcommand{\cS}{{\mathcal S}}

\newcommand{\cI}{{\mathcal I}}

\newcommand{\SU}{\mathrm{SU}}

\newcommand{\be}{\begin{equation}}
\newcommand{\ee}{\end{equation}}
\newcommand{\beq}{\begin{eqnarray}}
\newcommand{\eeq}{\end{eqnarray}}
\newcommand{\bes}{\begin{eqnarray}}
\newcommand{\ees}{\end{eqnarray}}

\newcommand{\su}{{\mathfrak{su}}}

\newcommand{\la}{\langle}
\newcommand{\ra}{\rangle}

\newcommand{\tr}{{\mathrm{Tr}}}
\newcommand{\f}{\frac}

\def\nn{\nonumber}
\def\pp{\partial}

\def\rd{\mathrm{d}}

\newcommand{\id}{\mathbb{I}}

\def\vphi{\varphi}
\def\eps{\epsilon}

\def\vJ{\vec{J}}

\def\tm{\tilde{m}}
\def\tbeta{\tilde{\beta}}
\def\talpha{\tilde{\alpha}}

\def\ta{\tilde{a}}

\newcommand{\alink}[4]
{\draw[decoration={markings,mark=at position 0.6 with {\arrow[scale=1.5,>=stealth]{>}}},postaction={decorate}] (#1) -- node[#3,pos=.5]{$#4$}(#2)}
\newcommand{\link}[2]
{\draw[decoration={markings,mark=at position 0.6 with {\arrow[scale=1.5,>=stealth]{>}}},postaction={decorate}] (#1) --(#2)}

\newtheorem{theorem}{Theorem}[section]

\newtheorem{proposition}[theorem]{Proposition}
\newtheorem{corollary}[theorem]{Corollary}
\newtheorem{definition}[theorem]{Definition}

\begin{document}

\title{Loop Quantum Gravity and Quantum Information}

\author{{\bf Eugenio Bianchi}}\email{ebianchi@psu.edu}
\affiliation{Institute for Gravitation and the Cosmos,  Pennsylvania State University, University Park, PA 16802, USA}
\author{{\bf Etera R. Livine}}\email{etera.livine@ens-lyon.fr}
\affiliation{Ecole Normale Sup\'erieure de Lyon,  Laboratoire de Physique (LP ENSL), CNRS (UMR 5672), 69007 Lyon,  France}

\date{\small February 12, 2023}

\begin{abstract}

\vspace*{2mm}

We summarize recent developments at the interface of quantum gravity and quantum information, and discuss applications to the quantum geometry of space in loop quantum gravity. In particular, we describe the notions of link entanglement, intertwiner entanglement, and boundary spin entanglement in a spin-network state. We discuss how these notions encode the gluing of quanta of space and their relevance for the reconstruction of a quantum geometry from a network of entanglement structures. We then focus on the geometric entanglement entropy of spin-network states at fixed spins, treated as a many-body system of quantum polyhedra, and discuss the hierarchy of volume-law, area-law and zero-law states. Using information theoretic bounds on the uncertainty of geometric observables and on their correlations, we identify area-law states as the corner of the Hilbert space that encodes a semiclassical geometry, and the geometric entanglement entropy as a probe of semiclassicality.

\vspace*{5mm}

{ Invited Chapter for the {\it Handbook of Quantum Gravity} (Eds. Bambi, Modesto and Shapiro, Springer 2023)}


\end{abstract}


\maketitle

\tableofcontents

\section{Introduction}

A physical and predictive description of gravitation at all scales of length and energy, or in short {\it quantum gravity}, is meant to emanate from consistently merging quantum theory and general relativity. In quantizing gravity or gravitizing the quantum, whichever perspective or starting point one might choose, a common thread is that we need to push further the revolutionary intuition of general relativity: gravity is not defined as a mere force or interaction but arises from the relations between observers; it is a manifestation of the relativity principle itself, applied to all observers and not restricted to inertial observers.

With geometry weaved from the network of relations between space-time points, it is then natural to seek a reformulation of gravity entirely in terms of information and flow. This point of view becomes even more relevant at the full quantum level. When the classical notions of coordinate system and reference frame fade away, the fundamental symmetry of general relativity --diffeomorphism invariance-- ineluctably leads to a localization problem. With no direct way to localize quantum systems in a quantum space-time without background geometry, only relations are physical degrees of freedom (see for instance \cite{Rovelli:1995fv} for such a reformulation of quantum mechanics). 
In this context, quantum geometry should be entirely defined in algebraic terms without referring to a classical background. Translating it to the language of quantum information would allow us to better understand its structure, in particular the flow from microscopic scales to macroscopic structures and the emergence of a classical smooth geometry.

Among the several  approaches to quantum gravity, {\it Loop Quantum Gravity} (LQG) builds space-time from Planck scale quanta of geometry. The fundamental degree of freedom is not the metric, but the Ashtekar-Barbero connection encoding the transport of reference frames along the space-time manifold. The metric is a composite field, distances are an emergent concept, and the quantum space-time is  built like an evolving lego system with quantum superpositions and quantum evolution of quantized reference frames.

In this context, we would like to use quantum information, and more specifically quantum entanglement, as a witness of quantum geometry, in order to probe the geometry at different scales and understand how classical geometry emerges.
In fact, a starting point is the natural hierarchy of correlations. For instance, 2-point correlation functions, in quantum field theories and condensed matter models, naturally allow to reconstruct distances between points, in suitable regimes. Then we expect 3-point correlations reveal corrections to the triangular inequality for distances and mismatch of transport around loops, i.e., to the notion of curvature. And so on, with multi-body entanglement revealing further layers of information about the geometry.

This chapter is thus devoted to introducing the basic mathematical notions to understand and study the structure of entanglement for  spin-network states of geometry in loop quantum gravity, and to providing useful examples illustrating the various sources of entanglement for quantum geometries.

\section{Quantum States of Geometry}

Loop quantum gravity is based on a Hamiltonian formulation of general relativity. At the classical level, we proceed to a 3+1 decomposition of space-time into a three-dimensional space and the one-dimensional time. Then general relativity predicts how the three-dimensional space manifold evolves in time. At the quantum level, loop quantum gravity defines quantum states of 3d geometry and describes their evolution in time. 

The quantum geometry is not directly described in terms of metric. As in other approaches to quantum gravity, the metric is a semi-classical notion emerging at large scales, in suitable regimes, from more fundamental quantum degrees of freedom. Loop quantum gravity (LQG) relies on the first order formulation of GR in terms of tetrad and Lorentz connection. The tetrad defines the local choice of basis for the four space-time dimensions, while the connection describes how to transport this local basis from a space-time point to another. The metric is a composite field reconstructed from both tetrad and connection. In those variables, classical general relativity is written as a gauge theory. The theory is invariant under both space-time diffeomorphisms and the Lorentz group as the local gauge group. We expect those symmetries to also drive the quantum theory.

LQG focuses on a complete set of observables: the holonomies (encoding the finite transport between space points defined by the connection) and fluxes (defining the normal bivectors to surfaces). The Poisson brackets between those variables, computed from the symplectic structure of general relativity, form a closed Lie algebra, called the holonomy-flux algebra. Its canonical quantization yields LQG and defines the unique diffeomorphism-covariant representation of those observables on quantum states (see \cite{Ashtekar:2004eh,Rovelli:2004tv,Thiemann:2007zz} for a thorough presentation and \cite{gambini2011first,rovelli2015covariant,Bodendorfer:2016uat,Ashtekar:2021kfp}  for a pedagogical overview).

Let us underline an important subtlety of the formalism. The fluxes and holonomies are canonically conjugate variables. However, the fluxes satisfy second class constraints, which reflect that they are entirely constructed from the tetrad fields even thought they have more components. The standard LQG formulation solves those ``simplicity'' constraints by choosing a specific gauge, called ``time gauge'', which sets the time-like normal vectors in internal space to the reference 4-vector $(1,0,0,0)$ \cite{BarberoG:1994eia,Immirzi:1996di} (see \cite{Samuel:2000ue,Alexandrov:2002br,Charles:2015rda} for issues related to this gauge fixing). This condition breaks the local Lorentz symmetry down to a local $\SU(2)$ symmetry under spatial rotations, so standard LQG is formulated in terms of $\SU(2)$ holonomies and $\SU(2)$ fluxes, as we will use in the present chapter. It is nevertheless to keep in mind that a modern framework has been developed which explicitly solves the simplicity constraints without gauge fixing and restores the Lorentz gauge symmetry \cite{Freidel:2020svx,Freidel:2020ayo}.

In this section, we define those quantum states of geometry in terms of spin networks. We recall their geometrical interpretation as discrete geometries and explain their re-interpretation as networks of entanglement between space-time points. In particular, we carefully distinguish the different types of degrees of freedom encoded in the spin network states and the entanglement they carry. This underlines the crucial role of two-dimensional boundaries in LQG and leads us to discuss the bulk-boundary relation and investigate holographic properties of spin networks.

\subsection{Wave-functions and Spin networks}\label{sec:spin-network}

\begin{figure}[h!]

\centering
\begin{tikzpicture}[scale=1.4]

\coordinate(a) at (0,0) ;
\coordinate(b) at (.5,1);
\coordinate(c) at (.9,-.1);
\coordinate(d) at (.3,-.8);
\coordinate(e) at (1.3,.7);
\coordinate(f) at (2.3,1.1);
\coordinate(g) at (2,.4);
\coordinate(h) at (2.7,.1);
\coordinate(i) at (2.1,-.2);
\coordinate(j) at (1.4,.-.5);

\draw (a) node {$\bullet$} node[left]{$s(e)$};
\draw (b) node {$\bullet$} node[above]{$t(e)$};
\draw (c) node {$\bullet$} ;
\draw (d) node {$\bullet$};
\draw (e) node {$\bullet$}node[above]{$v$};
\draw (f) node {$\bullet$};
\draw (g) node {$\bullet$};
\draw (h) node {$\bullet$} ++(-0.2,-0.7) node{$\Gamma$};
\draw (i) node {$\bullet$};
\draw (j) node {$\bullet$};

\alink{a}{b}{left}{g_{e}};
\link{a}{c};
\link{a}{d};
\link{b}{c};
\link{c}{d};
\link{b}{e};
\link{e}{f};
\link{c}{e};
\link{e}{g};
\link{f}{g};
\link{f}{h};
\link{h}{g};
\link{h}{i};
\link{g}{i};
\link{j}{i};
\link{d}{j};
\link{c}{j};
						
				
\end{tikzpicture}

\caption{Quantum states of geometry in loop quantum gravity are wave-functions of the $\SU(2)$ transport along the edges $e$ of a graph $\Gamma$. The  gauge invariance under local $\SU(2)$ transformations is imposed at every vertex $v$ of the graph.}
\label{fig:graph}
\end{figure}
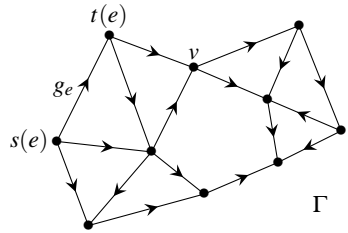

Loop quantum gravity defines quantum states that capture the geometry of a finite number of space-time degrees of freedom. Indeed, as illustrated on Fig.~\ref{fig:graph}, a state is defined by choosing a finite number of points on the 3d space manifold and a graph linking those points together and by considering a wave-function of the holonomies of the Ashtekar-Barbero connection along the links of the chosen graph. Those holonomies are group elements of the Lie group $\SU(2)$ and encode the transport, or change of 3d reference frame, from one point to another.
The emphasis on using the transport between reference frames as the fundamental variable for the geometry hardcodes relativity and the notion of change of observers in the kinematics of theory.

So let us consider a closed oriented graph $\Gamma$ with $E$ edges and $V$ vertices.
Focusing on a finite set of observables living on the graph --- the holonomies along the graph links do not necessarily mean that there is no geometry away from the chosen graph but that the whole geometry of space is to be entirely reconstructed from the data living on the graph \cite{Freidel:2011ue}. 
%
We consider gauge-invariant wave-functions depending on $\SU(2)$ group elements on every link or edge $e\in\Gamma$ while being invariant under the $\SU(2)$ action at every vertex $v\in\Gamma$:
\be
\label{eqn:closedwavefunction}
\vphi(\{g_{e}\}_{e\in\Gamma})=\vphi(\{h_{s(e)}g_{e}h_{t(e)}^{-1}\}_{e\in\Gamma})
\quad
\forall h_{v}\in\SU(2)^{\times V}
\,,
\ee
where $s(e)$ stands for the source vertex of the oriented $e$ and $t(e)$ the corresponding target vertex, as drawn on Fig.~\ref{fig:graph}. The symmetry requirement implements the gauge invariance under local $\SU(2)$ transformations, or equivalently local changes of 3d reference frame.

The Hilbert space of  quantum states of geometry on that graph is then the $L^{2}$ space of gauge-invariant wave-functions provided with the Haar measure on the Lie group $\SU(2)$:
\be
\cH_{\Gamma}=L^{2}(\SU(2)^{\times E}/\SU(2)^{\times V})\,,
\ee
\be
\la \vphi|\widetilde{\vphi}\ra_{\Gamma}
=\int_{\SU(2)^{{\times E}}}\prod_{e}\rd g_{e}\,
\overline{\vphi(\{g_{e}\}_{e\in\Gamma})}\,\widetilde{\vphi}(\{g_{e}\}_{e\in\Gamma})
\,.
\ee
The full Hilbert space of loop quantum gravity is then defined as the sum of those individual spaces $\cH_{\Gamma}$ over all graphs defined mathematically as a projective limit taking into account  the inclusion of graphs into one another \cite{Ashtekar:1994mh}.

\begin{figure}[h!]

\centering
\begin{tikzpicture}[scale=1.4]

\coordinate(a) at (0,0) ;
\coordinate(b) at (.5,1);
\coordinate(c) at (.9,-.1);
\coordinate(d) at (.3,-.8);
\coordinate(e) at (1.3,.7);
\coordinate(f) at (2.3,1.1);
\coordinate(g) at (2,.4);
\coordinate(h) at (2.7,.1);
\coordinate(i) at (2.1,-.2);
\coordinate(j) at (1.4,.-.5);

\draw (a) node {$\bullet$} node[left]{$I_{A}$};
\draw (b) node {$\bullet$}node[above]{$I_{B}$};
\draw (c) node {$\bullet$} node[right]{$I_{C}$};
\draw (d) node {$\bullet$} node[below]{$I_{D}$};
\draw (e) node {$\bullet$} node[above]{$I_{E}$};
\draw (f) node {$\bullet$};
\draw (g) node {$\bullet$};
\draw (h) node {$\bullet$};
\draw (i) node {$\bullet$};
\draw (j) node {$\bullet$};

\alink{a}{b}{left}{j_{1}};
\alink{a}{c}{above}{j_{2}};
\alink{d}{a}{left}{j_{3}};
\alink{b}{c}{left}{j_{6}};
\alink{c}{d}{right}{j_{4}};
\link{b}{e};
\link{e}{f};
\alink{c}{e}{right}{j_{5}};
\link{e}{g};
\link{f}{g};
\link{f}{h};
\link{h}{g};
\link{h}{i};
\link{g}{i};
\link{j}{i};
\link{d}{j};
\alink{c}{j}{right}{j_{7}};

\end{tikzpicture}

\caption{A spin network, on a closed oriented graph $\Gamma$, is a basis state is labeled with  spins $j_{e}\in\f N2$ on the graph edges and intertwiner states  $I_{v}$ on the graph vertices. Intertwiners are maps, that commute with the $SU(2)$ action, between the tensor product of incoming spins and the tensor product of outgoing spins. They can be understood as singlet states. Here, $I_{A}$ is a 3-valent intertwiner from $V^{j_{3}}$ to $V^{j_{1}}\otimes V^{j_{2}}$, while $I_{C}$ is a 5-valent intertwiner mapping $V^{j_{2}}\otimes V^{j_{6}}$ to $V^{j_{4}}\otimes V^{j_{5}}\otimes V^{j_{7}}$.
\label{fig:spinnetwork}}
\end{figure}
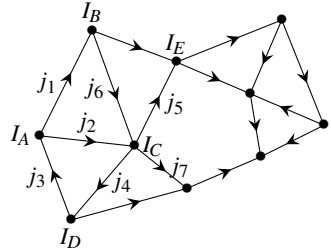

An orthonormal basis of  $\cH_{\Gamma}$ is defined by using the Peter-Weyl theoreom and the Plancherel decomposition of functions on $\SU(2)$. This is the equivalent of a Fourier decomposition, or more precisely the extension of the decomposition of functions on the two-sphere into spherical harmonics, where the Fourier modes are given by the Wigner matrices.
This leads the {\it spin network} basis states $|\Gamma,j_{e},I_{v}\ra$, labeled by a spin $j_{e}\in\f\N2$ on each edge $e$ and an intertwiner $I_{v}$ at every vertex $v$, as illustrated on Fig.~\ref{fig:spinnetwork}:
\be
\cH_{\Gamma}=\bigoplus_{\{j_{e},I_{v}\}_{e,v\in\Gamma}}\C\,|\Gamma,j_{e},I_{v}\ra\,.
\ee
A spin $j$ defines a irreducible representation of the Lie group $\SU(2)$. The associated  Hilbert space, noted $V^{j}$, has dimension $d_{j}=(2j+1)$. Writing $J^{a}$ with $a=x,y,z$ for the $\su(2)$ Lie algebra generators, we use the usual basis of $V^{j}$ diagonalizing both the $\su(2)$ Casimir $\vJ^{2}\equiv \delta_{ab}J^{a}J^{b}$ and the generator $J^{z}$, and labeled by the spin $j$ and the magnetic momentum $m$ running by integer step from $-j$ to $+j$:
\be
V^{j}=\bigoplus_{-j\le m\le j}\C\,|j,m\ra\,,
\quad
d_{j}=\dim V^{j}=2j+1
\,,
\ee
\be
\vJ^{2}\,|j,m\ra=j(j+1)\,|j,m\ra\,,\quad
J^{z}\,|j,m\ra=m\,|j,m\ra
\,.
\nn
\ee
An intertwiner $I_{v}$ at the vertex $v$ is a $\SU(2)$-invariant map between the tensor product of the incoming spins and the tensor product of the outgoing spins, as illustrated on Fig.~\ref{fig:spinnetwork}:
\be
I_{v}:\, \bigotimes_{e|t(e)=v} V^{j_{e}}\,\rightarrow \bigotimes_{e|s(e)=v} V^{j_{e}}
\,,\quad\textrm{such that}\quad
\forall h\in\SU(2)\,,\,\, h\circ I_{v}=I_{v}\circ h\,.
\ee
Since the complex conjugate representation $(V^{j})^{*}$ is isomorphic to $V^{j}$,  intertwiners can be simply seen as singlet states, in the tensor product of all the spins living on the edges linked to $v$:
\be
I_{v}\in \cI_{v}
=
\mathrm{Inv}_{\SU(2)}
\big{[}
\bigotimes_{e|s(e)=v}V^{j_{e}}\otimes\bigotimes_{e|t(e)=v}(V^{j_{e}})^{*}
\big{]}
\sim
\mathrm{Inv}_{\SU(2)}
\big{[}
\bigotimes_{e\ni v}V^{j_{e}}
\big{]}
\,.
\ee
By definition of an irreducible representation, bivalent intertwiners only exist if the two spins $j_{1}=j_{2}$ are equal and are then unique. Trivalent intertwiners between three spins exists if and only if those spins satisfy triangular inequalities, $|j_{1}-j_{2}|\le j_{3}\le(j_{1}+j_{2}) $ and are then unique: they are given by the Clebsh-Gordan coefficients. From valence 4 onwards, the intertwiner space grows in dimension and they are multiple non-trivial intertwiner states. An interesting result from the representation theory of semi-simple Lie algebras is that higher-valent intertwiners can always be made from 3-valent intertwiners by unfolding the vertex into a 3-valent tree.
%
%

Then the spin network wave-function is obtained by gluing the chosen intertwiners at the vertices together connecting them by the Wigner matrices of the holonomies along the graph edges:
\be
\vphi^{\{j_{e},I_{v}\}}\big{(}\{g_{e}\}\big{)}
=
\sum_{m_{e}^{s,t}}
\prod_{e}D^{j_{e}}_{m_{e}^{t}\,m_{e}^{s}}(g_{e})\,
\prod_{v}\la \bigotimes_{e|s(e)=v} j_{e}m_{e}^{s} |I_{v}| \bigotimes_{e|t(e)=v} j_{e}m_{e}^{t}\ra
\,,
\ee
where $D^{j}_{mm'}(g)=\la j,m|g|j,m' \ra$ are the matrix elements of  the Wigner matrix $D^{j}(g)$ representing the $\SU(2)$ group element $g$ in the spin-$j$ representation in the magnetic moment basis.

In the geometrical interpretation of loop quantum gravity \cite{Rovelli:1995ac,Ashtekar:1996eg,Ashtekar:1997fb}, intertwiners represent  excitations of the 3d volume while spins give the quanta of area of the interface gluing neighbouring chunks of volume (in other words, the cross-section). This is further validated by the geometrical interpretation of intertwiners as quantized convex polyhedra \cite{Freidel:2009ck,Bianchi:2010gc,Livine:2013tsa} with the spins giving the area of the polyhedra' faces. From this perspective, spin networks are interpreted as discrete  geometries, or more precisely the quantization of 3d twisted geometries \cite{Freidel:2010aq}. Twisted geometries are an extension of 3d Regge geometries to discontinuous 4d embedding \cite{Dittrich:2010ey}.

The goal of the present chapter is to introduce the tools to interpret the geometry of spin network in light of quantum information. In particular, we explain how the basic components of the quantum geometry -- the spins and intertwiners -- are reflected in the entanglement.

\subsection{Geometry as a Network of Entanglement}\label{sec:network-entanglement}

Now, LQG's quantum states of geometry are spin networks, built from gluing quanta of 3d volume --the intertwiners-- together through quanta of 2d area --the spin-network edges. This gluing is not an innocent operation; it actually creates entanglement between neighbouring quanta of 3d volume. So spin networks can be thought of as networks of entanglement.

In order to make this picture concrete, let us look more closely at the entanglement across a spin network edge, as depicted in Fig.~\ref{fig:edge}.
There are three different notions of entanglement, depending of the objects on which we focus:
\begin{itemize}
\item the {\it link entanglement} between the two spins that are directly glued together by the link;

\item the {\it intertwiner entanglement} between the two intertwiners living at the vertices glued by the link;

\item the {\it boundary spin entanglement} between the other spins around those vertices.

\end{itemize}
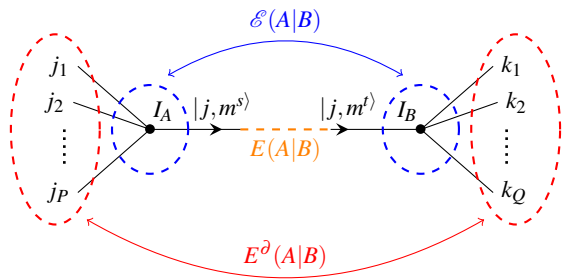
\begin{figure}[h]
\begin{center}

%
%
%
%
%
%
%
%
%
\begin{tikzpicture}[scale=1.2]

\coordinate(a) at (0,0) ;
\coordinate(b) at (3,0);
\coordinate(ah) at (1,0) ;
\coordinate(bh) at (2,0);

\draw (a) node {$\bullet$};
\draw (b) node {$\bullet$};
\draw (a) node {$\bullet$} ++(0.15,0.2) node{$I_{A}$};
\draw (b) node {$\bullet$} ++(-0.15,0.2)  node{$I_{B}$};
\draw[dashed,thick,orange] (ah)-- node[midway,below]{$E(A|B)$} (bh);
\draw[decoration={markings,mark=at position 0.8 with {\arrow[scale=1.5,>=stealth]{>}}},postaction={decorate}] (a) -- node[above,pos=.8]{$|j,m^{s\ra}$}(ah);
\draw[decoration={markings,mark=at position 0.2 with {\arrow[scale=1.5,>=stealth]{>}}},postaction={decorate}] (bh) -- node[above,pos=.2]{$|j,m^{t\ra}$}(b);

\draw (a)--++(-.85,0.3) node[left]{$j_{2}$};
\draw (a)--++(-.8,0.7) node[left]{$j_{1}$};
\draw (a)--++(-.8,-.7)  node[left]{$j_{P}$};

\draw (b)--++(.85,0.3) node[right]{$k_{2}$};
\draw (b)--++(.8,0.7) node[right]{$k_{1}$};
\draw (b)--++(.8,-.7)  node[right]{$k_{Q}$};

\draw[dotted,line width=1pt] (-.95,0) -- (-.95,-.4);
\draw[dotted,line width=1pt] (3+.95,0) -- (3+.95,-.4);

\draw[dashed, thick,blue] (0,0) ellipse (12pt and 14pt);
\draw[dashed, thick,blue] (3,0) ellipse (12pt and 14pt);
\draw[<->,blue] (.2,.6)  to[bend left] node[midway,above,blue]{$\cE(A|B)$} (2.8,.6);

\draw[dashed, red, thick] (-1.05,0) ellipse (14pt and 30pt);
\draw[dashed,red, thick] (4.05,0) ellipse (14pt and 30pt);
\draw[<->,red] (-.7,-1)  to[bend right] node[midway,above,red]{$E^{\pp}(A|B)$} (3.7,-1);

\end{tikzpicture}

\caption{We distinguish the {\color{orange} link entanglement  $E(A|B)$} between the spin states living at the two ends of the spin network edge, the {\color{blue} intertwiner entanglement  $\cE(A|B)$}  between the two intertwiners living at the vertices $A$ and $B$, and the  {\color{red}  boundary spin entanglement $E^{\pp}(A|B)$} between the spins living on the boundary of the spin network region formed by merging the two  3d regions $A$ and $B$.
\label{fig:edge}}

\end{center}
\end{figure}

Let us underline that spin states $|j,m\ra$ live on the half-edges of the spin networks. They are not $\SU(2)$-gauge invariant object and they are in fact summed over in the definition of the spin network wave-function. From that point of view, the link entanglement does not measure the quantum entanglement between physical degrees of freedom.
On the other hand, intertwiners are legitimate gauge-invariant objects and reveal quantum correlations between the two spin network vertices. However, they do not reflect at all whether an actual link exists between those two vertices or not.
Nevertheless, putting these two notions together leads to the boundary spin entanglement. Indeed, as we will prove below, the boundary spin entanglement is indeed the sum of the link entanglement and intertwiner entanglement. 

While glued spin states along spin network edges are not physical degrees of freedom, spin states living on cut links have a different status. Indeed, they are interpreted as living on the  boundary of the spin network. Here, considering two vertices glued by one edge, the spins living on all the other links attached to the vertices constitute the boundary of that region of space. They become legitimate degrees of freedom (see e.g. \cite{Donnelly:2016auv} for a reformulation of boundary modes as would-be-gauge degrees of freedom in gauge theories, relevant for LQG) and the boundary spin entanglement is a tangible measure of the correlations between them.

\subsubsection{Link entanglement}

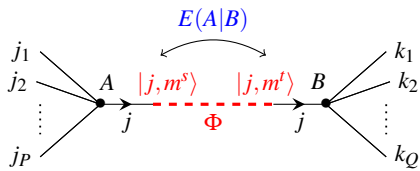
\begin{figure}[h]
\begin{center}

\begin{tikzpicture}[scale=1]

\coordinate(a) at (0,0) ;
\coordinate(b) at (3,0);
\coordinate(ah) at (.7,0) ;
\coordinate(bh) at (2.3,0);

\draw (a) node {$\bullet$} ++(0.1,0.3) node{$A$};
\draw (b) node {$\bullet$} ++(-0.1,0.3)  node{$B$};
\alink{a}{ah}{below}{j};
\alink{bh}{b}{below}{j};
\draw[dashed,very thick,red] (ah)--node[midway, below]{$\Phi$} (bh) ;

\draw (ah)++(0.2,0) node[above,red]{$|j,m^{s}\ra$};
\draw (bh)++(-0.1,0) node[above,red]{$|j,m^{t}\ra$};

\draw (a)--++(-.85,0.3) ;
\draw (a)--++(-.8,0.7);
\draw (a)--++(-.8,-.7);

\draw (b)--++(.85,0.3) ;
\draw (b)--++(.8,0.7) ;
\draw (b)--++(.8,-.7)  ;

\draw (a)--++(-.85,0.3) node[left]{$j_{2}$};
\draw (a)--++(-.8,0.7) node[left]{$j_{1}$};
\draw (a)--++(-.8,-.7)  node[left]{$j_{P}$};

\draw (b)--++(.85,0.3) node[right]{$k_{2}$};
\draw (b)--++(.8,0.7) node[right]{$k_{1}$};
\draw (b)--++(.8,-.7)  node[right]{$k_{Q}$};

\draw[dotted,line width=.7pt] (-.8,0) -- (-.8,-.4);
\draw[dotted,line width=.7pt] (3+.8,0) -- (3+.8,-.4);

\draw[<->] (.8,.65)  to[bend left] node[midway,above,blue]{$E(A|B)$} (2.2,.65);

\end{tikzpicture}

\caption{The two spin network vertices $A$ and $B$ are glued along a link carrying the spin $j$. The link entanglement  {\color{blue} $E(A|B)$ }along that spin network edge is the entanglement between the source spin {\color{red} $|j,m^{s}\ra$} and the target spin {\color{red} $|j,m^{t}\ra$} defined by the chosen  link state $\Phi$, which controls the gluing. This link state links in $V^{j}\otimes (V^{j})^{*}$ and represents the superposition of $\SU(2)$ holonomies transproting the local reference frame  from $A$ to $B$.
\label{fig:link}}

\end{center}
\end{figure}

As drawn on Fig.~\ref{fig:link}, a spin network link $e$ carries two spin states, one on each end of the link.
These two spin states are related by the holonomy $g\in\SU(2)$ running along the link.
Let us assume that the link is dressed with a fixed spin $j$. Then the state associated to the link lives in $V^{j}\otimes (V^{j})^{*}= \textrm{End}[V^{j}]$.  In the case that the holonomy is known, then the state is:
\be
\Phi^{(j,g)}
=
\f{1}{\sqrt{d_{j}}}\sum_{m^{s},m^{t}}
D^{j}_{m^{t}m^{s}}(g)\,|j,m^{t}\ra \la j,m^{s}|
\in V^{j}\otimes (V^{j})^{*}
\,.
\ee
Since the Wigner matrices are unitary, this state\footnotemark{} is normalized.
\footnotetext{Using  the ismorphism between $V^{j}$ and $(V^{j})^{*}$, we can map this into a state in $V^{j}\otimes V^{j}$:
\be
\Phi_{e}^{g} \mapsto 
\f{1}{\sqrt{d_{j}}}\sum_{m^{s},m^{t}} (-1)^{j-m^{s}}
D^{j}_{m^{t}m^{s}}(g)\,|j,m^{t}\ra \otimes | j,-m^{s}\ra
\in 
V^{j}\otimes V^{j}
\,.
\ee
In the special case where the holonomy is trivial along the link, $g=\id$, then the Wigner matrix is the identity and the link state reduces to the standard singlet state $\sum_{m} (-1)^{j-m}\,|j,m\ra\otimes  |j,-m\ra\
\in\,\mathrm{Inv}_{\SU(2)}\big{[}
V^j\otimes V^j\big{]}$.
}
This clearly is a maximally entangled state between the source and target. To check this, we compute the reduced density matrix, for instance for the target spin, which simplifies due the unitarity of the Wigner matrix:
\be
\rho_{t}^{(j,g)}=\f1{d_{j}}\sum_{m^{t}}|j,m^{t}\ra\la j,m^{t}|=\id_{j}\,,
\ee
where $\id_{j}$ is the identity on the Hilbert space $V^{j}$.
This reduced density matrix does not depend on the holonomy $g$ and is always the totally mixed state, with maximal entropy:
\be
E^{(j,g)}(s|t)=-\tr \rho_{t}^{(j,g)}\ln\rho_{t}^{(j,g)} =\ln d_{j}=\ln(2j+1)\,.
\ee
The link entanglement is non-zero as soon as the spin carried by the link does not vanish $j \ne 0$. In fact, considering a vanishing spin $j=0$ is completely equivalent to having no link. Thus having a non-trivial link between two vertices in a spin network means having a non-trivial entanglement.  
This leads to the picture of a spin network as a graph whose links represent a (maximal) entanglement between spin states, or in simpler terms, an entanglement graphs  \cite{Colafranceschi:2020ern,Colafranceschi:2022ual}.

Nevertheless, in general, the holonomy and the spins are not fixed. They are distributed according to the considered graph wave-function. This is appropriately described by a link state that depends on a probability amplitude for the holonomy $\phi(g)\in L^{2}(\SU(2))$. By decomposing this probability amplitude into spins by the Peter-Weyl theorem, one translates it into the corresponding link state:
\be
\phi(g)=\sum_{j,m^{s},m^{t}}\sqrt{d_{j}}\phi^{j}_{m^{t}m^{s}}D^{j}_{m^{t}m^{s}}(g)
\,\,\mapsto\,\,
\Phi=\sum_{j,m^{s},m^{t}}\phi^{j}_{m^{t}m^{s}}\,|j,m^{t}\ra\la j,m^{s}|\,
\ee
\be
\textrm{with the normalization}\quad
\int \rd g\,|\phi(g)|^{2}=1=\sum_{j}\tr_{j}\,\phi^{j}(\phi^{j})^{\dagger}=\la \Phi|\Phi\ra
\,.
\ee
The reduced density matrix for the target spin reads:
\be
\rho_{t}^{\phi}=\tr_{s}|\Phi\ra\la \Phi|
=
\sum_{j,m^{t}\tm^{t}}\left[\phi^{j}(\phi^{j})^{\dagger}\right]_{m^{t}\tm^{t}}\,|j,m^{t}\ra\la j,\tm^{t}|
\,\,\in\,\textrm{End}\Big{[}\bigoplus_{j}V^{j}\Big{]}
\,.
\ee
This matrix is diagonal by block, each block corresponding to a spin $j$, and its Von Neumann entropy gives  the entanglement carried by the link:
\be
E_{\Phi}(s|t)=-\sum_{j}\tr_{}(\rho_{j}\ln\rho_{j})\quad
\textrm{with}\,\,\,
\rho_{j}=\phi^{j}(\phi^{j})^{\dagger}
\,.
\ee
One can fix the spin by requiring that the probability amplitude $\phi$ for the holonomy carries a single spin $j_{0}$, i.e. that the  matrices $\phi_{j}$ vanish except for $j=j_{0}$. In this case, the entanglement sum formula reduces to a single term:
\be
\phi_{j}\propto \delta_{jj_{0}}
\quad\Rightarrow\quad
E_{\Phi}(s|t)=-\tr_{}(\rho_{j_{0}}\ln\rho_{j_{0}})\quad
\textrm{with}\,\,\,
\tr\,\rho_{j_{0}}=1
\,.
\ee
Then the entanglement is bounded by the value of the spin $E_{\Phi}(s|t)\le \ln(2j_{0}+1)$. The maximal value is reached when $\rho_{j_{0}}$ is the totally mixed state, in which case we recover the result obtained earlier for fixed spin and holonomy:
\be
\rho_{j}=\delta_{jj_{0}}\,\f1{d_{j_{0}}}\id_{j_{0}}\quad \Rightarrow\quad E_{\Phi}(s|t)=\ln(2j_{0}+1)\,.
\ee
This means that, for a fixed spin, non-maximal entanglement along a spin network link comes from quantum superpositions of holonomies.

\subsubsection{Intertwiner entanglement vs. Boundary spin entanglement}

\begin{figure}[h]
\begin{center}

\begin{tikzpicture}[scale=1.5]

\coordinate(a) at (0,0) ;
\coordinate(b) at (1.5,0);

\draw (a) node {$\bullet$} ++(0.1,0.15) node{$I_{A}$};
\draw (b) node {$\bullet$} ++(-0.1,0.15)  node{$I_{B}$};
\alink{a}{b}{below}{j};

\draw (a)--++(-.85,0.3) ;
\draw (a)--++(-.8,0.7);
\draw (a)--++(-.8,-.7);

\draw (b)--++(.85,0.3) ;
\draw (b)--++(.8,0.7) ;
\draw (b)--++(.8,-.7)  ;

%

\draw[dotted,line width=.7pt] (-.8,0) -- (-.8,-.4);
\draw[dotted,line width=.7pt] (1.5+.8,0) -- (1.5+.8,-.4);

\draw[dashed, blue,thick] (0,0.1) ellipse (12pt and 12pt);
\draw[dashed, blue,thick] (1.5,0.1) ellipse (12pt and 12pt);
\draw[<->] (.2,.8)  to[bend left] node[midway,above]{$\cE(A|B)$} (1.3,.8);

\end{tikzpicture}
\hspace*{10mm}
\begin{tikzpicture}[scale=1]

\coordinate(a) at (0,0) ;
\coordinate(b) at (1.5,0);

\draw (a) node {$\bullet$};
\draw (b) node {$\bullet$};
\alink{a}{b}{above}{j};

\draw (a)--++(-.85,0.3) node[left]{$j_{2}$};
\draw (a)--++(-.8,0.7) node[left]{$j_{1}$};
\draw (a)--++(-.8,-.7)  node[left]{$j_{P}$};

\draw (b)--++(.85,0.3) node[right]{$k_{2}$};
\draw (b)--++(.8,0.7) node[right]{$k_{1}$};
\draw (b)--++(.8,-.7)  node[right]{$k_{Q}$};

\draw[dotted,line width=1pt] (-.95,0) -- (-.95,-.4);
\draw[dotted,line width=1pt] (1.5+.95,0) -- (1.5+.95,-.4);

\draw[dashed,red, thick] (-1.05,0) ellipse (16pt and 30pt);
\draw[dashed,red, thick] (2.55,0) ellipse (16pt and 30pt);
\draw[<->] (-.7,1)  to[bend left] node[midway,above]{$E_{\pp}(A|B)$} (2.2,1);

\end{tikzpicture}

\caption{Intertwiner entanglement entropy (on the left) versus boundary spin state entanglement (on the right): we distinguish the entanglement $\cE(A|B)$ between the intertwiner states at the two nodes in {\color{blue} \bf blue} and  the entanglement $E_{\pp}(A|B)$ they induce on the boundary spins in {\color{red} \bf red}.
\label{fig:2vertex}}

\end{center}
\end{figure}
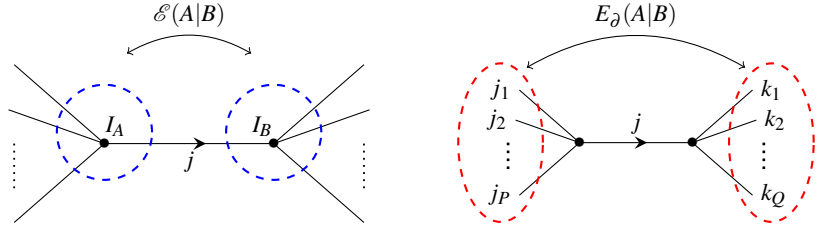

The link entanglement along a spin network edge only reflects the act of gluing two vertices along that edge. It does not provide a measure of the quantum correlations between the quanta of 3d geometry living at the two vertices linked by that edge. Let us thus focus on two neighboring vertices of a spin network, $A$ and $B$, as on the Fig.~\ref{fig:2vertex}, linked by a single edge oriented from $A$ to $B$ and decorated with a fixed given spin $j$. The other legs attached to the vertex $A$ are decorated with spins $j_{1},..,j_{P}$, while the other edges attached to $B$ carry the spins $k_{1},..,k_{Q}$.
Having fixed all the spins, the only remaining freedom is the choice of intertwiner states. Indeed the Hilbert space of  spin network states for this configuration at fixed spins is:
\be
\cH_{AB}=\cI_{A}\otimes\cI_{B}\,,
\ee
where $\cI_{A}$ and $\cI_{B}$ are the spaces of intertwiners attached to the two nodes:
\be
\cI_{A}=\mathrm{Inv}_{\SU(2)}
\big{[}
V^{j_{1}}\otimes..\otimes V^{j_{P}}
\otimes (V^{j})^{*}
\big{]}
\,,\,\,\,\,
\cI_{B}=\mathrm{Inv}_{\SU(2)}
\big{[}
V^{k_{1}}\otimes..\otimes V^{k_{Q}}
\otimes V^{j}
\big{]}
\,.
\ee
We assumed that all the boundary spins $j_{p}$ and $k_{q}$ are incoming onto $A$ and $B$.

Considering a pure state in $\cH_{AB}$, we define the intertwiner entanglement $\cE(A|B)$ between $A$ and $B$ as the Von Neumann entropy of the reduced density matrices, obtained by tracing over $\cI_{A}$ or $\cI_{B}$.
Let's be more explicit by choosing an orthonormal basis for the intertwiners at the two nodes:
\be
I_{A}^\alpha\in\cI_{A}=\mathrm{Inv}\big{[}
\bigotimes_{p=1}^PV^{j_{p}}\otimes (V^j)^{*}
\big{]}
\,,
\quad
\la I_{A}^{\alpha}\,|\,I_{A}^{\talpha}\ra=\delta_{\alpha\talpha}\,,
\ee
\be
I_{B}^\beta\in\cI_{B}=\mathrm{Inv}\big{[}
\bigotimes_{q=1}^QV^{k_{q}}\otimes V^j
\big{]}
\,,
\quad
\la I_{B}^{\beta}\,|\,I_{B}^{\tbeta}\ra=\delta_{\beta\tbeta}\,,
\ee
For an arbitrary normalized spin network state $I\,\in\cH_{AB}=\cI_{A}\otimes\cI_{B}$,
\be
|I\ra=\sum_{\alpha\beta}\psi_{\alpha\beta}\,|I_{A}^\alpha\ra\otimes |I_{B}^\beta\ra
\,,\quad
\la I\,|\,I\ra=\tr \psi\psi^\dagger=1
\,,
\ee
we compute the reduced density matrix:
\be
\rho_{A}
=
\tr_{B}\,|I\ra\la I|
=
\sum_{\alpha,\talpha}
\big{(}\psi\psi^\dagger\big{)}_{\alpha\talpha}\,
|I_{A}^\alpha\ra\la I_{A}^{\talpha}|
\,,
\quad
\tr\rho_{A}=1
\,,
\ee
from which we get the intertwiner entanglement:
\be
\cE_{I}(A|B)=-\tr\,\psi\psi^\dagger\ln\,\psi\psi^\dagger
\,.
\ee
Note that, even though the coefficient matrix $\psi$ is a priori not a square matrix, the matrix $\psi\psi^\dagger$ is always a square matrix. 
This entanglement entropy is bounded by the dimension of the intertwiner spaces:
\be
\cE_{I}(A|B)\le \min\big{(}
\ln \dim\cI_{A},
\ln \dim\cI_{B}
\big{)}\,.
\ee

Another way to look at the regions $A$ and $B$, once glued by the intermediate edge, is to group them into a single region $AB$. The edge between $A$ and $B$, carrying the spin $j$, is now within this coarser region $AB$, while the remaining edges carrying the spins $j_{1},..,j_{P},k_{1},..,k_{Q}$ are all on its boundary. Then the Hilbert space of boundary spin states is:
\be
H^{\pp}_{AB}=
\underbrace{\big{(}
V^{j_{1}}\otimes..\otimes V^{j_{P}}
\big{)}}_{H_{A}^{\setminus (AB)}}
\otimes
\underbrace{\big{(}
V^{k_{1}}\otimes..\otimes V^{k_{Q}}
\big{)}}_{H_{B}^{\setminus (AB)}}
\,,
\ee
which is clearly not the same Hilbert space as $\cH_{AB}$.
We write $H_{A}^{\setminus (AB)}=\bigotimes_{p}V^{j_{p}}$ for the Hilbert space tensor product of all the spins attached to the vertex $A$ but the one living on the edge $(AB)$ linking the two vertices. Similarly for $B$.
The natural bipartite splitting of this boundary Hilbert space leads to a notion of boundary spin  entanglement, a priori different from the intertwiner entanglement. As shown in \cite{Livine:2017fgq}, these two notions are nevertheless related.
In order to make this explicit, we write the boundary spin states obtained from a spin network configuration by gluing the two vertices $A$ and $B$ by a link state of determined spin $j$:

\begin{definition}- {\bf Gluing map from Intertwiners to Boundary spins}
\vspace*{1mm}\\
{\it For a factorized spin network state $I_{A}\otimes I_{B}\in\cH_{AB}$ with decoupled intertwiner states at the two vertices $A$ and $B$,  and a normalized link state $\Phi$ of fixed spin $j$,
\be
\Phi=\sum_{a,b}\phi^{j}_{ba}\,|j,b\ra\la j,a|\quad\in\,\,V^{j}\otimes (V^{j})^{*}
\,,\quad
\la\Phi|\Phi\ra=\tr\phi^{j}(\phi^{j})^{\dagger}=1
\,,
\ee
we define the corresponding boundary spin state in $H^{\pp}_{AB}$ obtained by gluing the two intertwiners through this link state as
\be
\cP_{\Phi}[I_{A}\otimes I_{B}]
=
\la \Phi |I_{A}\otimes I_{B}\ra_{j\otimes j^{*}}
=
d_{j}\sum_{a,b}\phi^{j}_{ba}I_{A}^{a}\otimes I_{B}^{b}
\quad\in
H^{\pp}_{AB}\,,
\ee
where $d_{j}=(2j+1)$ is a normalization factor.
The notations $I_{A}^{a}$ and $I_{B}^{b}$ correspond to the projection of the intertwiner states on fixed magnetic moment of the spin $j$ component:
\be
I_{A}=\sum_{a}I_{A}^{a}\otimes \la j,a|\in \cI_{A}
\quad\textrm{with}\,\,I_{A}^{a}\in H_{A}^{\setminus (AB)}
\ee
\be
I_{B}=\sum_{b}I_{B}^{b}\otimes | j,b\ra\in \cI_{B}
\quad\textrm{with}\,\,I_{B}^{b}\in H_{B}^{\setminus (AB)}
\ee
The map $\cP_{\Phi}$ is then extended to the whole Hilbert space $\cI_{A}\otimes \cI_{B}$ by linearity.}
\end{definition}

A special case is when the holonomy along the linking edge is fixed  $g\in\SU(2)$. The resulting gluing map $\cP_{g}$ then reads:
\be
\Phi_{g}=\f1{\sqrt{d_{j}}}\sum_{a,b}D^{j}_{ba}(g)\,|j,b\ra\la j,a|
\,,
\ee
\be
\cP_{g}:\,I_{A}\otimes I_{B}\in\cH_{AB} 
\,\mapsto\,
{\sqrt{d_{j}}}\sum_{a,b}D^{j}_{ba}(g)
I_{A}^{a} \otimes I_{B}^{b}
\quad\in \,H^{\pp}_{AB}
\,.
\ee
The trivial  gluing corresponds to  $g=\id$, for which $\cP_{\id}[I_{A}\otimes I_{B}]\,={\sqrt{d_{j}}}\sum_{a}
I_{A}^{a} \otimes I_{B}^{a}$.

\begin{proposition}
\label{prop:superposition}
- {\bf Entanglement of Intertwiner Superpositions:}
\vspace*{1mm}\\
Let $I\in\cH_{AB}=\cI_{A}\otimes \cI_{B}$ be an arbitrary normalized  superposition of tensor product of intertwiner states.
Then the corresponding boundary state $\cP_{\Phi}\,[I] \in H^{\pp}_{AB}$, obtained by gluing the  intertwiners along a link state of fixed spin-$j$ carries an entanglement simply equal to the sum of the intertwiner entanglement of $I$ and of the link entanglement of $\Phi$:
\be
E^{\pp}_{\cP_{\Phi}\,[I]}(A|B)=\cE_{I}(A|B)+E_{\Phi}(A|B)\,.
\ee

\end{proposition}
\begin{proof}
For an arbitrary normalized spin network state $I\,\in\cH_{AB}=\cI_{A}\otimes\cI_{B}$,
\be
|I\ra=\sum_{\alpha\beta}\psi_{\alpha\beta}\,|I_{A}^\alpha\ra\otimes |I_{B}^\beta\ra
\,,\quad
\la I\,|\,I\ra=\tr \psi\psi^\dagger=1
\,,
\ee
and a normalized link state of fixed spin $j$,
\be
\Phi=\sum_{a,b}\phi^{j}_{ba}\,|j,b\ra\la j,a|\quad\in\,\,V^{j}\otimes (V^{j})^{*}
\,,\quad
\la\Phi|\Phi\ra=\tr\phi^{j}(\phi^{j})^{\dagger}=1
\,,
\ee
the boundary spin state resulting from gluing  the intertwiners at  $A$ and $B$  using the link state $\Phi$s
\be
\cP_{\Phi}[I]
=
d_{j}\sum_{\alpha\beta}\psi_{\alpha\beta}\,\cP_{\Phi}[I_{A}^\alpha\otimes I_{B}^\beta]
=
d_{j}\sum_{\alpha\beta}\sum_{a,b}\psi_{\alpha\beta}\,\phi^{j}_{ba}I_{A}^{\alpha,a}\otimes I_{B}^{\beta,b}
\,,
\ee
with the notation $I_{A}^{\alpha,a}=(I_{A}^{\alpha})^{a}\in \,H_{A}^{\setminus (AB)}=\bigotimes_{p=1}^PV^{j_{p}}$. The key property of  these intertwiner components is that they are orthonormal:
\be
\la I_{A}^{\alpha,a} | I_{A}^{\talpha,\ta}\ra
=\,
\f1{d_{j}}\,
\delta_{\alpha,\talpha}\delta_{a,\ta}
\,.
\ee
Indeed, each intertwiner basis label $a$ defines an embedding of the tensor product $\bigotimes_{p=1}^PV^{j_{p}}$ into the single spin space $V^j$, i.e. a channel recoupling the spins $j_{1},..,j_{P}$ into the spin $j$. This implies that projecting the intertwiner on different magnetic moments $a$ lead to orthogonal states in $V^j\hookrightarrow\bigotimes_{i=p}^PV^{j_{p}}$.
%
%
The same holds for the node $B$.
This allows to check explicitly that the boundary spin state  is normalized, $\la\, \cP_{\Phi}[I]\,|\,\cP_{\Phi}[I]\,\ra=1$.
We can then compute the reduced density on the boundary spins of $A$, as an endomorphism of $H_{A}^{\setminus (AB)}=\bigotimes_{i=p}^PV^{j_{p}}$:
\be
\rho^{\pp}_{A}=
\sum_{\alpha\talpha}\sum_{a,\ta}
(\psi\psi^\dagger)_{\alpha\talpha}
\, (\phi^{j\dagger}\phi^{j})_{\ta a}
\,
d_{j}|I_{A}^{\alpha,a}\ra\la I_{A}^{\talpha,\ta}|
\,,\quad
\tr\,\rho^{\pp}_{A}=1
\,.
\ee
This directly gives the spin state entanglement:
\be
E^{\pp}_{\cP_{\Phi}[I]}(A|B)
=
-\big{(}\tr\,\phi^\dagger\phi\ln\,\phi^\dagger\phi\big{)}-\big{(}\tr\,\psi\psi^\dagger\ln\,\psi\psi^\dagger\big{)}
=
\cE_{I}(A|B)
+
E_{\Phi}(A|B)
\,.
\ee
thus concluding the proof of this proposition extending the results of \cite{Livine:2017fgq}.

\end{proof}

A direct consequence of this proposition is its application to pure tensor product states, which do not carry any intertwiner entanglement, $E(A|B)=0$, in which case the  boundary spin entanglement $E^{\pp}(A|B)$ exactly reduces to the link entanglement.  
In particular:
%
\begin{corollary}
- {\bf Factorized Intertwiner States:}
\vspace*{1mm}\\
Let $I_{A}\otimes I_{B}\in\cH_{AB}=\cI_{A}\otimes \cI_{B}$ be a normalized intertwiner tensor product state. The resulting boundary spin entanglement resulting from gluing $A$ and $B$ with and edge with fixed $j$ and given holonomy $g\in\SU(2)$ only depends on the spin $j$ and does not depend on the holonomy $g$:
\be
E^{\pp}_{\cP_{g}[I_{A}\otimes I_{B}]}(A|B)=\ln(2j+1)\,.
\ee
\end{corollary}
This result is the insight that the spins of the spin network reflect the quantum correlation between spin network vertices, i.e. quanta of 3d volumes as underlined in \cite{Livine:2006xk,Donnelly:2008vx,Donnelly:2011hn}. It is in fact at the origin of the proposal of relating the spins to a notion of distance in loop quantum gravity \cite{Livine:2006xk}.

An interesting feature is that the entanglement above does not depend on the holonomy between the two vertices. This is consistent with the local $\SU(2)$ gauge invariance of the theory: one can always change the $\SU(2)$ holonomy along the edge linking the two nodes $A$ and $B$, and in particular set $g_{AB}=\id$, by doing a suitable $\SU(2)$ gauge transformation at $A$ or at $B$. This is actually at the heart of the ``coarse-graining by gauge-fixing'' procedure introduced in \cite{Livine:2013gna,Charles:2016xwc}. Now, if one was to consider three nodes, $A$, $B$ and $C$, forming a loop in the spin network graph, one could not gauge-away the $\SU(2)$ holonomy around the closed loop $A\rightarrow B\rightarrow C\rightarrow A$ and the tri-partite entanglement between the three intertwiners will depend non-trivially on this holonomy and reflect the curvature carried by the spin network \cite{Chen:2022rty}.

\subsection{Boundary Spins and Bulk Reconstruction}\label{sec:boundary-bulk}


The boundary spin point of view for spin network is much more general than the analysis of neighbouring vertices.
Below, we extend the definition of the boundary spin entanglements to bounded regions of a spin network containing more than two vertices.
It allows for a formulation of spin network states of quantum geometry for 3d regions with 2d boundaries and opens the door to a systematic study of the relation between bulk geometry and boundary states.

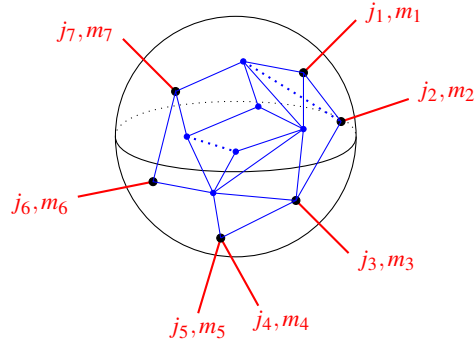
\begin{figure}[!htb]
\centering
\begin{tikzpicture}
\coordinate(O) at (0,0);
\draw (O) circle (1.6);
\draw[dotted,in=90,out=90,looseness=.5] (-1.6,0) to (1.6,0);
\draw[in=-90,out=-90,looseness=.5] (-1.6,0) to (1.6,0) ++(0,-.4);
\coordinate (b1) at (.9,.85);
\coordinate (b2) at (1.4,.2);
\coordinate (b3) at (.8,-.85);
\coordinate (b45) at (-.2,-1.35);
\coordinate (b6) at (-1.1,-0.6);
\coordinate (b7) at (-.8,.6);

\draw[red,thick](b1)--++(.7,.7) ++(0,0.1)node[right,red]{$j_{1},m_{1}$};
\draw[red,thick](b2)--++(.9,.3) ++(0,0.1)node[right,red]{$j_{2},m_{2}$};
\draw[red,thick](b3)--++(.7,-.7) ++(0,-0.1)node[right,red]{$j_{3},m_{3}$};
\draw[red,thick](b45)--++(.5,-.9) ++(-0.2,-0.2)node[right,red]{$j_{4},m_{4}$};
\draw[red,thick](b45)--++(-.3,-1) ++(0.5,-0.2)node[left,red]{$j_{5},m_{5}$};
\draw[red,thick](b6)--++(-1,-.2) ++(0,-0.1)node[left,red]{$j_{6},m_{6}$};
\draw[red,thick](b7)--++(-.7,.7) ++(0,.1)node[left,red]{$j_{7},m_{7}$};

\filldraw[black] (b1) circle (1.5pt);
\filldraw[black] (b2) circle (1.5pt);
\filldraw[black] (b3) circle (1.5pt);
\filldraw[black] (b45) circle (1.5pt);
\filldraw[black] (b6) circle (1.5pt);
\filldraw[black] (b7) circle (1.5pt);

\coordinate (v1) at (0.1,1);
\filldraw[blue] (v1) circle (1pt);
\coordinate (v2) at (0.3,.4);
\filldraw[blue] (v2) circle (1pt);
\coordinate (v3) at (0.9,.1);
\filldraw[blue] (v3) circle (1pt);
\coordinate (v4) at (-.65,0);
\filldraw[blue] (v4) circle (1pt);
\coordinate (v5) at (0,-.2);
\filldraw[blue] (v5) circle (1pt);
\coordinate (v6) at (-0.3,-.75);
\filldraw[blue] (v6) circle (1pt);

\draw[blue] (b7)--(v1)--(b1)--(b2)--(b3)--(b45);
\draw[blue,thick,dotted] (v1)--(b2);
\draw[blue] (v1)--(v2)--(v3)--(v1);
\draw[blue] (b1)--(v3)--(b3);
\draw[blue] (v2)--(v4)--(v6)--(v5)--(v3);
\draw[blue,thick,dotted] (v4)--(v5);
\draw[blue] (b3)--(v6)--(v3);
\draw[blue] (b45)--(v6)--(b6)--(b7)--(v4);

\end{tikzpicture}
	\caption{Loop quantum gravity boundary data defined by the spin network puncturing the two-dimensional space boundary $\cS_{2d}=\pp \Sigma_{3d}$: the interior grapg $\gamma$ is drawn in {\color{blue} \bf blue}; the spin network boundary edges $e\in\pp\Gamma$ are drawn in {\color{red}\bf red}, they connect to the interior graph $\gamma$  by the boundary vertices in {\bf black} and carry boundary spin states $|j_{i},m_{i}\ra$, which are the quantization of the geometrical flux through the boundary surface and define quanta of area on $\cS_{2d}$.}
	\label{fig:boundedregion}
\end{figure}

Indeed, the standard definition of loop quantum gravity's quantum states as wave-functions on closed graphs, as given earlier in \eqref{eqn:closedwavefunction}, is only valid for closed 3d space manifolds, without boundaries. Now, a two-dimensional boundary in 3d space would cut links of the graph and delimit a bounded region of a spin network state, as depicted on Fig.~\ref{fig:boundedregion}.  The geometry of a region with 2d boundary should thus be described with states living on open graphs.

Let us call $\Gamma$ an open graph. And let us distinguish the boundary $\pp\Gamma$, as the set of open edges, from the interior graph $\gamma$ made from all the edges whose both ends are contained in the graph, $\gamma=\Gamma \setminus\pp\Gamma$.

The boundary describes the boundary of the 3d region. Each open edge carries a spin state with an a priori arbitrary spin. This defines the boundary Hilbert space associated to a single boundary edge, or in other words to a single puncture on the 2d boundary:
\be
\forall e\in\pp\Gamma\,,\quad
\cH_{e}^{\pp} =\bigoplus_{j_{e}\in\f\N2}V^{j_{e}}\,.
\ee
The boundary Hilbert space is then the  tensor product of those single spin spaces:
\be
\cH_{\pp\Gamma}=\bigotimes_{e\in\pp\Gamma}\cH_{e}^{\pp}
\,.
\ee
A boundary state in $\cH_{\pp\Gamma}$  describes the quanta of areas of the boundary surface and can be thought of as a quantum boundary condition for the bulk geometry.

Quantum states of the bulk geometry are now defined as {\it gauge-covariant} wave-functions of the $\SU(2)$ holonomies along the edges of the interior graph $\gamma$, valued in the boundary Hilbert space \cite{Chen:2021vrc}:
\be
\psi_{\Gamma}:\SU(2)^{E_{\gamma}}\rightarrow \cH_{\pp\Gamma}\,.
\ee
This means that, for a set of group elements $\{g_{e}\}_{e\in\gamma}$ on the graph, the evaluation of the wave-function $\psi_{\Gamma}[\{g_{e}\}]$ is a boundary state. Then, for a given boundary state $\Omega\in \cH_{\pp\Gamma}$, the squared modulus of the scalar product in boundary space,
\be
P^{\Omega}_{\psi}[\{g_{e}\}]
=
\big{|}\la \Omega\,|\,\psi_{\Gamma}[\{g_{e}\}]\ra_{\pp}\big{|}^{2}\,,
\ee
gives the probability for the bulk holonomies  $\{g_{e}\}$ for the fixed quantum boundary condition $\Omega$.
%
%
We do not require the wave-function to be fully invariant under local $\SU(2)$ gauge transformations at each vertex of the graph, but to be $\SU(2)$-covariant under boundary gauge transformations.
This is a generic feature of boundaries in gauge field theories (see e.g. \cite{Donnelly:2016auv}).
More precisely, in our framework, we distinguish boundary vertices, to which are attached the boundary edges, from bulk vertices, which are not connected to any boundary edge. Then gauge-invariance translates into:
\be
\psi_{\Gamma}[\{h_{t(e)}g_{e}h_{s(e)}^{-1}\}]
=
\bigotimes_{e\in\pp\Gamma} h_{v(e)}^{\eps_{e}}
\,\psi_{\Gamma}[\{g_{e}\}]
\quad\in\,\,\cH_{\pp\Gamma}
\,,
\ee
where the vertex $v(e)$ denotes the (boundary) vertex to which the boundary edge $e\in\pp\Gamma$ is attached, and where $\eps_{e}=1$ when the boundary edge is outgoing (i.e. $v(e)=s(e)$) and $\eps_{e}=-1$ when the boundary edge is incoming (i.e. $v(e)=t(e)$).
When the boundary is empty, $\pp\Gamma=\emptyset$, then the boundary Hilbert space is trivial, $\cH_{\pp}=\C$ and we recover standard wave-functions and spin networks on closed graphs.

This formulation of bulk quantum states as linear maps on the boundary Hilbert space, or in short ``boundary maps'', as introduced in \cite{Chen:2021vrc}, allows for a clearer bulk-boundary relation.
For instance, it is expected that tracing out the bulk degrees of freedom, one obtains a mixed boundary state \cite{Livine:2006xk,Bianchi:2013toa},  which can be seen to lead to a bulk-induced decoherence mechanism for the boundary \cite{Feller:2016zuk, Livine:2017xww}. 
Here, the boundary state induced by the bulk state $\psi_{\Gamma}$ is simply given by the boundary density matrix defined as:
\be
\rho_{\pp}[\psi_{\Gamma}]=\int_{\SU(2)^{E_{\gamma}}} \prod_{e}\rd g_{e}\,
|\psi_{\Gamma}[\{g_{e}\}]\ra\la \psi_{\Gamma}[\{g_{e}\}]|
\quad\in\,\,\,\textrm{End}\Big{[} \cH_{\pp\Gamma}\Big{]}\,.
\ee
This is the general formulation to describe the classical and quantum correlations between boundary spins, extending the previous definition of boundary spin entanglement reviewed in the previous section \ref{sec:network-entanglement}.

This density matrix actually encodes the state of the boundary once we trace out the bulk geometry and encodes all the correlations and entanglement carried by the boundary due to its embedding in the quantum geometry defined by the bulk state  $\psi_{\Gamma}$. For instance, in the case of a black hole described in loop quantum gravity, this density matrix would contain the entanglement information between parts of a black hole horizon, as considered for example in \cite{Livine:2005mw}.

\medskip

The natural question one may ask, especially in the context of the research on the holographic principle in quantum gravity, is: how much can we know on the bulk state $\psi_{\Gamma}$, and in particular on the graph $\Gamma$, from the boundary density matrix $\rho_{\pp}[\psi_{\Gamma}]$?
As such, the {\it bulk reconstruction} problem is reformulated as the {\it purification} of the boundary state.
A universal reconstruction theorem was shown in \cite{Livine:2017xww,Chen:2021vrc}: whatever the  density matrix on the boundary Hilbert space, one can always induce it from a spin network states based on a graph with a single loop in the bulk. The key is that the $\SU(2)$ holonomy around that loop is enough freedom to explore the whole space of boundary density matrices. Since the holonomies carried by the spin networks represent the curvature in the bulk, this theorem reflects the insight that  the entanglement of the boundary geometry of the region reflects the quantum fluctuations of the curvature within the region (see \cite{Chen:2022rty} for further developments).  However, it also shows that the boundary data does not reflect the depth of the region. Thus a truly holographic reconstruction of the bulk from the boundary crucially depends on the properties of the spin network states and requires a deeper understanding of the entanglement carried by the bulk states. 
Setting the foundations of the structure of entanglement in LQG is the focus of the next section.

\section{Geometric Entanglement Entropy and the Hierarchy of States}

The kinematical Hilbert space of loop quantum gravity is spanned by spin-network states $|\Gamma, j_e, I_v\rangle$, an orthonormal basis of states that, at fixed graph $\Gamma$, simultaneously diagonalizes a complete set of commuting observables given by ultra-local operators which have the geometric interpretation of microscopic areas of each link and microscopic volume of each node of the graph. As a result, by construction, correlations at space-like separation vanish: For instance the connected volume-volume correlation function is identically zero in each spin-network basis state. The following two questions then naturally arises: What is the behavior of correlation functions in a \emph{typical} state in the Hilbert space? And how do we characterize the \emph{corner} of the Hilbert space that supports states with long-range correlations as the ones of perturbative quantum fields in a fixed background spacetime? In this section we delineate recent progress in addressing these two questions using quantum-information methods.

\subsection{Entanglement Entropy Bounds on Uncertainties and Correlations}\label{sec:entropy-bounds}
The notion of entanglement entropy provides a powerful tool that allows us to characterize properties of a subsystem, such as uncertainties and correlations, in terms of a single information-theoretic quantity. We describe here two results and their application to the quantum geometry of space in loop quantum gravity.

\medskip

The first key result is a lower bound on the uncertainty of outcomes of measurements on a subsystem. Let us consider an isolated quantum system with bipartite Hilbert space $\mathcal{H}=\mathcal{H}_A\otimes\mathcal{H}_B$. Given a pure state $|\psi\rangle$, we can determine the probability $p_k$ that in a measurement of an observable $\mathcal{O}_A$ in the subsystem $A$ we find as outcome the eigenvalue $\lambda_k$. A standard way of characterizing the probability distribution of measurement outcomes is in terms of moments of the distribution: We compute the average outcome $\langle\lambda\rangle$, the dispersion around the average $\Delta \lambda$ and, if the probability distribution is peaked around the average, these two moments characterize well its peakedness properties. In general, however, for distributions with long tails and for multimodal distributions, average and dispersion do not provide us with a quantitative characterization of the uncertainty in the distribution of measurement outcomes. Information theoretic methods, such us the Shannon entropy of a probability distribution $S(p_k)=-\sum_k p_k \log p_k$, provide us exactly with the tool to address questions about uncertainty in these more general cases: a large Shannon entropy indicates large uncertainty in the distribution of measurement outcomes, while a zero Shannon entropy tells us that the state $|\psi\rangle$ is an eigenstate of the observable $\mathcal{O}_A$, and therefore there is a certain outcome given by the associated eigenvalue.
The entanglement entropy $S_A(|\psi\rangle)=-\mathrm{Tr}_A(\rho_A\log\rho_A)$ (with $\rho_A=\mathrm{Tr}_B|\psi\rangle\langle\psi|$) provides a bound on the entropy of measurement for any set of observables in the subsystem $A$. Let us consider a state $|\psi\rangle$ and a complete set of commuting observables  $\mathcal{O}_A^i$ (c.s.c.o.) in the subsystem $A$. The entropy of measurement outcomes is 
\begin{equation}
S(|\psi\rangle,\;\mathcal{O}_A^i\rightarrow \alpha^i)=-\sum_{\alpha_i} p(|\psi\rangle,\;\mathcal{O}_A^i\rightarrow \alpha^i)\,\log p(|\psi\rangle,\;\mathcal{O}_A^i\rightarrow \alpha^i)\,,
\label{eq:measurement-entropy}
\end{equation}
where the probability distribution is given by the Born rule, $p_k=\sum_{\beta}|\langle \alpha^i,\beta|\psi\rangle|^2$, and $|\alpha^i,\beta\rangle=|\alpha^i\rangle_A\otimes|\beta\rangle_B$ is an orthonormal basis of eigenstates of the c.s.c.o. $\mathcal{O}_A^i$. This quantity is bounded from below by the entanglement entropy \cite{nielsen00}:
\begin{equation}
S_A(|\psi\rangle)\;\leq\; S(|\psi\rangle,\;\mathcal{O}_A^i\rightarrow \alpha^i)\qquad \forall\;  \mathcal{O}_A^i\; \;\textrm{c.s.c.o. in } \mathcal{H}_A\,.
\label{eq:uncertainty-bound}
\end{equation}
Note that, while the right-hand side of the inequality depends both on the state and on the observables, the entanglement entropy on the left-hand side depends only on the state and on the subsystem $A$ that those observables probe. Therefore the entanglement entropy provides a universal \emph{lower bound on uncertainties} for observables in a subsystem. In the next section we use this result to put bounds on the uncertainty of geometric observables in loop quantum gravity.

\medskip

The second key result is an upper bound on correlations of observables. Let us consider an isolated quantum system with tripartite Hilbert space $\mathcal{H}=\mathcal{H}_A\otimes\mathcal{H}_B\otimes\mathcal{H}_C$. Given a pure state $|\psi\rangle$, we can compute the connected correlation function $\mathcal{G}$ of two bounded observables $\mathcal{O}_A$ and $\mathcal{O}_B$ acting respectively on subsystems $A$ and $B$:
\begin{equation}
\mathcal{G}\;=\;\langle \psi|\mathcal{O}_A\,\mathcal{O}_B|\psi\rangle-\langle \psi| \mathcal{O}_A|\psi\rangle\,\langle \psi|\mathcal{O}_B|\psi\rangle\,.
\label{eq:correlation-function}
\end{equation}
The mutual information $S_{AB|C}(|\psi\rangle)$ of the subsystems $A$ and $B$ can be expressed in terms of the entanglement entropies of $A$, $B$ and $C$ and is given by
\begin{equation}
S_{AB|C}(|\psi\rangle)=S_A(|\psi\rangle)+S_B(|\psi\rangle)-S_C(|\psi\rangle)\,.
\label{eq:mutual-information}
\end{equation}
Remarkably, the mutual information provides us with an upper bound on the correlation function $\mathcal{G}$, \cite{nielsen00,Wolf}:
\begin{equation}
\frac{\big(\langle \psi|\mathcal{O}_A\,\mathcal{O}_B|\psi\rangle-\langle \psi| \mathcal{O}_A|\psi\rangle\,\langle \psi|\mathcal{O}_B|\psi\rangle\big){}^2}{2\,\|\mathcal{O}_A\|^2 \|\mathcal{O}_B\|^2}\leq S_{AB|C}(|\psi\rangle)\,.
\label{eq:correlations-bound}
\end{equation}
Note that, while the left-hand side of the inequality depends on the observables $\mathcal{O}_A$ and $\mathcal{O}_B$, the right-hand side depends only on the subspaces they act on. Therefore the mutual information provides a universal \emph{upper bound on correlations}. In the next section we use this result to put bounds on  correlations of geometric observables in loop quantum gravity.

\medskip

Throughout the analysis we assume that the system is in a pure state $|\psi\rangle$. As a result, the entropy in the complement $\mathcal{H}_C$ equals the entropy in the factor $\mathcal{H}_A\otimes\mathcal{H}_B$ probed by the set of observables $\mathcal{O}_A$ and $\mathcal{O}_B$, i.e.,  $S_C(|\psi\rangle)=S_{AB}(|\psi\rangle)$ in (\ref{eq:mutual-information}). We note that the assumption that the system is in a pure state is not a restriction: We adopt the point of view that, if the quantum geometry of space was in a mixed state, then we simply have to take into account a larger Hilbert space that includes the purifying degrees of freedom. For instance, in the presence of matter, the complement $\mathcal{H}_C$ describes both the quantum geometry not probed by the observables $\mathcal{O}_A$ and $\mathcal{O}_B$, and the matter degrees of freedom.

\subsection{Hierarchy of States: Volume-Law, Area-Law and Zero-Law States}\label{sec:hierarchy}
In loop quantum gravity, the quantum geometry of space is described by a state $|\psi\rangle$ in the kinematical Hilbert space $\mathcal{H}^{\mathrm{kin}}$. A semiclassical state is peaked on a specific value of the intrinsic and the extrinsic geometry of space. This condition fixes only the expectation value and dispersion of local geometric operators, but leaves correlations at space-like separations unspecified. On the other hand, in perturbative quantum field theory on a given classical background geometry, these correlations take a specific long-range form. It becomes crucial then to identify the class of semiclassical states $|\psi\rangle$ that capture the perturbative effective-field-theory regime of the theory, by prescribing also correlation functions. Quantum information methods, and in particular the results (\ref{eq:uncertainty-bound}) and (\ref{eq:correlations-bound}), provide a new perspective on how to restrict the class of semiclassical states by moving the focus from correlation functions to a hierarchy of scaling laws for the entanglement entropy of a region: volume-law, area-law and zero-law states. As we will see, area-law states provide a condition for semiclassicality.

\medskip

The kinematical Hilbert space $\mathcal{H}^{\mathrm{kin}}$ is spanned by spin-network states $|\Gamma,j_e,I_v\rangle$ with graph $\Gamma$, spin $j_e$ associated to its edges and intertwiners $I_v$ associated to its vertices.  To illustrate the main ingredients of the hierarchy of scaling laws, we consider a specific sector of $\mathcal{H}^{\mathrm{kin}}$: We fix both the graph $\Gamma$ and the spins $j_e$. As a result, we obtain the sector
\begin{equation}
\mathcal{H}^{\mathrm{kin}}_{\Gamma, j_\ell}=\bigotimes_{v\in \Gamma} \mathcal{I}_v(j_e)
\label{eq:product-H}
\end{equation}
where $\mathcal{I}_v(j_e)$ is the intertwiner Hilbert space at the vertex $v$. Having fixed the graph $\Gamma$, we have an immediate notion of connectivity and of geometric regions. Moreover, at fixed spins, the Hilbert space reduces to a tensor product over finite dimensional Hilbert spaces $\mathcal{I}_v$, one per node. This is the same structure present in ordinary many-body quantum system, and we can easily bring in methods and perspectives from condensed matter theory. We discuss later how to generalize these methods to a sum over spins and graphs.

For concreteness, we restrict attention to a regular graph: a cubic lattice $\Gamma$ with a finite number $N$ of sites (or vertices) and periodic boundary conditions (torus topology). We assume also that the spins associated to each edge of the graph are fixed and equal to $j_0$. As a result, we have a many-body quantum system, where each body is a quantum polyhedron \cite{Bianchi:2010gc} with six faces of equal area $a_0=8\pi G \hbar \gamma \sqrt{j_0(j_0+1)}$. We can write the generic state as
\begin{equation}
|\psi\rangle=\sum_{i_1,\ldots,i_N}\psi_{i_1,\ldots,i_N}\;|i_1\rangle\otimes\cdots\otimes |i_N\rangle\,,
\label{eq:generic-state}
\end{equation}
with $|i_n\rangle \in \mathcal{I}_v=\textrm{Inv}_{SU(2)}[(V^{j_0}){}^{\otimes 6}]$ an orthornormal basis of intertwiners at each vertex. The factorized structure (\ref{eq:product-H}) allows us to identify an immediate notion of geometric regions containing $N_A$ contiguous vertices, and geometric entanglement entropy associated to the subsystem decomposition $\mathcal{H}=\mathcal{H}_A\otimes\mathcal{H}_B$. We are now ready to describe a hierarchy of states with different properties of the quantum geometry.

\medskip

\emph{Zero-Law States}. Spin-network basis states $|\Gamma,j_0,i_n\rangle=|i_1\rangle\otimes\cdots\otimes |i_N\rangle$ are product states over the intertwiners associated to the vertices of the cubic lattice. As a result, if we consider a region $A$ of the lattice, the entanglement entropy $S_A$ vanishes: they are zero-law states. To illustrate their properties, let us consider some examples. We can take $|i_n\rangle$ to be given by eigenstates of the oriented volume operator $V_n$. In this case, dihedral angles $\theta_{ne\tilde{e}}$ will have large uncertainties (with a non-zero probability of finding both cuboids and pentagonal wedges), but the outcomes of measurements of the volume have zero measurement entropy, $S(|\psi\rangle,V_n\rightarrow v_n)=0$. Therefore they saturates the uncertainty bound (\ref{eq:uncertainty-bound}). We can also consider coherent states for the intrinsic geometry of a cubic lattice by choosing the state (\ref{eq:generic-state}) with $\psi_{i_1,\ldots,i_N}=\Phi_{i_1}\cdots \Phi_{i_N}$ and $\Phi_{i_n}$ coherent intertwiners peaked over the Euclidean geometry of a cube \cite{Freidel:2010aq,Bianchi:2010gc,Livine:2013tsa}. In this case, the average volume is the one of a cube with faces of fixed area, the average dihedral angles are right angles, but there are fluctuations around the average. Are fluctuations at nearby nodes correlated? The correlation bound (\ref{eq:correlations-bound}) tells us that all connected correlation functions, such as angle-angle correlations $\mathcal{G}=\langle \theta_{ne\tilde{e}}\,\theta_{n'e'\tilde{e}'}\rangle-\langle \theta_{ne\tilde{e}}\rangle\langle\theta_{n'e'\tilde{e}'}\rangle\;=0$, vanish because the mutual information between regions is zero. Therefore there are no correlations at space-like separation for zero-law states.

\medskip

\emph{Volume-Law States}. The Hilbert space of a cubic lattice at fixed spin $j_0$ has finite dimension $\mathrm{dim}\mathcal{H}_{\Gamma,j_0}=(\mathrm{dim}\mathcal{I}_n)^N$ where $N$ is the number of nodes in the lattice and  $\textrm{dim}\mathcal{I}_n$ is the dimension of each intertwiner space. For a finite dimensional Hilbert space, there is a notion of random state $|\psi\rangle$, that is a state extracted randomly from the ensemble of normalized vectors with uniform probability distribution. Equivalently, one can consider a reference state $|\psi_0\rangle$ and a random state $|\psi\rangle=U|\psi_0\rangle$, where $U$ is a random unitary extracted from the ensemble of unitary matrices distributed according to the Haar measure. Remarkably, the entanglement entropy of random states has typicality properties: For a random state, the probability of finding a given value of the entanglement entropy is peaked at an average value $\langle S_A\rangle$ with a small dispersion $\Delta S_A$  \cite{page1993average,Bianchi:2019stn,Bianchi:2021aui}. Let us consider a region $A$ that contains a finite fraction of the nodes of the cubic lattice, and discuss the limit of large region $N_A\to \infty$, and large lattice, $N\to \infty$, with fixed finite ratio $f_A$:
\begin{equation}
f_A=\tfrac{N_A}{N}\leq \tfrac{1}{2}\,.
\end{equation}
In a random state, the average entanglement entropy of the region $A$ is
\begin{equation}
\langle S_A\rangle\;=\; f_A V\,\tfrac{\log d_n}{v_0} \;-\tfrac{1}{2}d_n{}^{-(1-2f_A)V/v_0}\;+\;O\big(d_n{}^{-V/v_0}\big)\,,
\label{eq:average-S}
\end{equation}
where $v_0$ is the average volume of each node. The number $d_n=\textrm{dim}\mathcal{I}_n$ is the dimension of each $6$-valent intertwiner space. It is equal to $d_n=5$ for $j_0=1/2$, and scales as $d_n\sim \frac{8}{\pi}j_0{}^3$ for $j_0\gg 1$. As the formula shows, at the leading order, the entanglement entropy of the region $A$ scales linearly with the volume $V_A=f_A V$ of the region. While this is the average entanglement entropy of random states, its value is also \emph{typical}: the dispersion around the average is exponentially small in the volume, 
\begin{equation}
\Delta S_A\;=\;\sqrt{\tfrac{1}{2}-\tfrac{1}{4}\delta_{f,1/2} }\;\;d_n{}^{-(1-f_A)V/v_0}\;+\;O(d_n{}^{-V/v_0})\,.
\label{eq:delta-S}
\end{equation}
The uncertainty bound (\ref{eq:uncertainty-bound}) tells us that, because of the typical volume law for a random state, there is no observable in the region $A$ that has vanishing uncertainty. In fact for a set of observables probing a region of average volume $V_A$, i.e., a ${}\;$c.s.c.o. in $\mathcal{H}_A$, the measurement entropy (\ref{eq:measurement-entropy}) scales linearly with the volume. Moreover, one can consider the mutual information for two regions $A$ and $B$. The leading order volume-law terms cancel in (\ref{eq:mutual-information})  and, as a result, the typical mutual information in a random state is exponentially small in the total  volume $V$. We can use then the correlations bound (\ref{eq:correlations-bound}) to conclude that, for instance,  angle-angle correlations in a random state scale as
\begin{equation}
\mathcal{G}\;=\;\langle \theta_{ne\tilde{e}}\,\theta_{n'e'\tilde{e}'}\rangle-\langle \theta_{ne\tilde{e}}\rangle\langle\theta_{n'e'\tilde{e}'}\rangle\;\;=\;\;O\big(d_n{}^{-V/2v_0}\big)\,, 
\end{equation}
and in general, correlation functions between any two space-like separated observables are exponentially small in a random state.

\medskip

\emph{Area-Law States}. In many-body quantum systems, area-law states arise as low-energy states of Hamiltonians with local interactions. To lower the average energy, the state needs to have a long-range decay of correlation functions which result in geometric entanglement entropy that scales with the area of the boundary of the region \cite{bombelli1986quantum,srednicki1993entropy,eisert2010colloquium}. This behavior is to be contrasted to the one of high-energy states which typically satisfy a volume law. In fact, random states of fixed energy (away from the edge of the energy spectrum) show a thermal behavior for subsystems (known as Eigenstate Thermalization Hypothesis) \cite{deutsch1991,srednicki1994,rigol.2008} and result in a typical volume-law entanglement entropy with scaling coefficient dependent on the energy \cite{Bianchi:2019stn,Bianchi:2021aui}. 
Another relevant example of states that do not belong to the low-energy spectrum is zero-law states, which  by definition are product states: the lack of space-like correlations results in a contribution to the spatial coupling in the Hamiltonian which makes them high-energy states. In quantum field theory, these zero-law states do not even belong to the Fock space of the theory as they are not Hadamard states \cite{Birrell:1982ix,wald1994quantum} and, if an ultraviolet cutoff is introduced, they can be understood as states with divergent energy density. To better characterize the class of area-law states in a way that is insensitive to the ultraviolet behavior of the theory, it is useful to consider the mutual information $S_{AB|C}(|\Psi\rangle)$ between a region $A$ and the complement $B$ of the enlarged region $AC$, where $C$ is a ``safety corridor'' around $A$ \cite{Casini:2008wt}. In loop quantum gravity, area-law states at fixed spins arise from long-range intertwiner correlations. Their geometric interpretation is discussed in the next section. 

We note here that the conjectured relation between entanglement in loop quantum gravity and entanglement in quantum field theory relies on the same key assumption used in the investigation of the graviton propagator in spinfoams \cite{Bianchi:2006uf,Bianchi:2011hp}. In quantum field theory, there is a fixed background geometry that allows us to choose a region and then compute entanglement of quantum fields. In loop quantum gravity, when one considers a semiclassical state, the expectation value of the geometry determines an effective background and the quantum fluctuations of the geometry reproduce the correlations of the metric perturbations in the effective field theory. It is this matching of correlation functions that results in a matching of entanglement entropies in the semiclassical regime.

\subsection{Area-law states and the architecture of spacetime geometry}\label{sec:architecture}
In a non-perturbative theory of quantum gravity, low-energy-density arguments (as the ones discussed above for many-body systems and quantum field theory) are problematic as they assume the existence of a classical background geometry with respect to which the energy density is defined. It is then useful to reverse the logic. The architecture conjecture \cite{Bianchi:2012ev} uses quantum-information methods as a probe of semiclassicality of physical states in quantum gravity:
\begin{itemize}[leftmargin=1em,noitemsep]
\item[] \emph{Entanglement and the architecture of space-time}.  In a theory of quantum gravity, for any sufficiently large causal domain in a semiclassical space-time, the entanglement entropy between the degrees of freedom describing a given causal domain $R$ and those describing its complement is finite and, to leading order, takes the area-law form $S_R(|\psi\rangle)=2\pi\, \mathrm{Area}(\partial R)/L_P^2\,+\ldots\,$, where $|\psi\rangle$ is the quantum state of the semiclassical space-time geometry and $\mathrm{Area}(\partial R)$ is the expectation value of the area of the $2$-dimensional corner of the causal domain $R$.
\end{itemize}
The conjecture is motivated by the area-law behavior of the vacuum state in many-body systems with local interactions, and by the vacuum entanglement entropy in quantum fields theory on Minkowski and on curved spacetimes. In quantum field theory, the vacuum entanglement entropy is ultraviolet divergent and scales with the area only if an ultraviolet cut-off is introduced, either by putting the theory on a lattice \cite{bombelli1986quantum,srednicki1993entropy}, or by introducing a safety corridor $C$ \cite{Casini:2008wt}, or by defining the region using a coarse-grained sub-algebra of observables \cite{Bianchi:2019pvv}. On the other hand, in quantum gravity the finite Planck area coefficient $L_P^2/2\pi$ is expected, as supported by various lines of evidence from the renormalization of the gravitational coupling in black hole backgrounds \cite{Susskind:1994sm}, from the thermodynamics of space-time geometry \cite{Jacobson:1995ab}, from the holographic entanglement entropy \cite{Ryu:2006bv}, and from the entanglement entropy of the Rindler horizon in perturbative quantum gravity \cite{Bianchi:2013rya}.

Often, it is useful to characterize states in the Hilbert space independently of the Hamiltonian and directly in terms of properties of the correlation functions. This procedure identifies a \emph{corner} of the Hilbert space that can then be used as a variational ansatz for the dynamics. In many-body quantum systems, this approach has a computational advantage over exact diagonalization of the Hamiltonian. Using a similar strategy in loop quantum gravity, where a local notion of energy is not available, allows us to identify classes of states with a desired scaling of the correlation functions before addressing the difficult problem of the dynamics. A general technique for addressing this problem is provided by \emph{squeezed vacua} for spin-networks, a family of states $|\Gamma,\gamma\rangle$ that spans the Hilbert space of loop quantum gravity, which are labeled by a graph $\Gamma$ and a matrix $\gamma_{ij}^{AB}$ which encodes quantum correlations  \cite{Bianchi:2016hmk,Bianchi:2016tmw,Bianchi:2015fra}. Squeezed vacua provide an overcomplete basis of loop quantum gravity that is tailored to the study of the entanglement structure of space in a semiclassical spacetime geometry with perturbative quantum fluctuations. The definition of squeezed vacua for loop quantum gravity involves three key ingredients. The first is the use of a bosonic Hilbert space as done in the spinor representation of loop quantum gravity \cite{Girelli:2005ii,Borja:2010rc,Livine:2011gp,Livine:2013wmq}. The idea is based on Schwinger's oscillator model of spin \cite{schwinger:1952osc}: Given a two oscillators with creation operators $a_1^\dagger$, $a_2^\dagger$ and vacuum $|0\rangle$, the state of definite spin is given by $\textstyle |j,m\rangle=\tfrac{(a_1^\dagger)^{j+m}}{\sqrt{(j+m)!}}\tfrac{(a_2^\dagger)^{j-m}}{\sqrt{(j-m)!}}|0\rangle$. The second ingredient is the construction of an overcomplete basis of squeezed vacua for a bosonic lattice introduced in \cite{Bianchi:2015fra}. In the simple case of a single harmonic oscillator with vacuum $|0\rangle$, a  squeezed vacuum is given by $|\gamma\rangle=e^{\,\frac{1}{2} \gamma\, a^\dagger a^\dagger}|0\rangle\,$, where $\gamma$ is a complex number. The third ingredient is an improvement of the original loop expansion that is at the roots of loop quantum gravity \cite{Rovelli:1989za}. Using bosonic variables, one can introduce normal-ordered Wilson loops and use the loop expansion to define a projector from bosonic states to loop states \cite{Bianchi:2016hmk}.  
The result of this construction is a formulation of squeezed vacua for loop quantum gravity, written as a superposition of spin-network basis states $|\Gamma, j_e, I_\nu\rangle$, (Eq.~102 in \cite{Bianchi:2016hmk}):
\begin{equation}
|\Gamma, \gamma^{\,AB}_{ij}\rangle=\sum_{j_e, I_\nu} c_{j_e, I_\nu}(\gamma^{\,AB}_{ij})\,|\Gamma,j_e, I_\nu\rangle
\end{equation}
where the expansion coefficients $c_{j_e, i_n}(\gamma^{\,AB}_{ij})\equiv \langle \Gamma,j_e, I_\nu|\Gamma, \gamma^{\,AB}_{ij}\rangle$ are expressed as 
 Gaussian integrals
\begin{equation}
c_{j_e, I_\nu}(\gamma^{\,AB}_{ij})=\int\tfrac{d^{4E}z\,d^{4E}\bar{z}}{\pi^{4E}}\;Z_{j_e, I_\nu}\,\mathrm{e}^{-z_i^A\bar{z}^i_A+\frac{1}{2}\gamma^{\,AB}_{ij}\bar{z}^i_A\bar{z}^j_B}\,,
\label{eq:c-coeff}
\end{equation}
with the polynomial insertion
\begin{equation}
Z_{j_e, I_\nu}=\sum_{m_i=-j_i}^{+j_i}
\Big(
\prod_{\nu=1}^{N} [\bar{I}_\nu]_{m_1\cdots m_{|\nu|}}
\Big)\;
\Big(
\prod_{i=1}^{2E} 
\tfrac{(z^0_i)^{j_i-m_i}}{\sqrt{(j_i-m_i)!}}
\tfrac{(z^1_i)^{j_i+m_i}}{\sqrt{(j_i+m_i)!}}
\Big)\,.
\end{equation}  
The advantage of using squeezed vacua is that they encode efficiently properties of a physically-relevant corner of the Hilbert space into a matrix $\gamma_{ij}^{\,AB}$, which determines the correlations at space-like separation that can be bound using formula (\ref{eq:correlations-bound}). 
Note that the matrix $\gamma^{\,AB}_{ij}$ appears only in the exponent in (\ref{eq:c-coeff}) and it couples the spinorial variables $z^A_i$ attached to the endpoints of the links of the graph $\Gamma$. We note also that squeezed vacua automatically induce a superposition over spins with specific coefficients which include also $j_e=0$ and therefore a sum over all subgraphs. Information-theoretic methods allow us to characterize quantum states of the spacetime geometry in terms of their entanglement structure and mutual information. In particular, for $\gamma^{\,AB}_{ij}=0$, the state reduces to the Ashtekar-Lewandowski vacuum (projected to the graph $\Gamma$) and it satisfies a zero-law for the geometric entanglement entropy. A non-trivial area law arises for Bell-network states \cite{Baytas:2018wjd,Bianchi:2018fmq} which corresponds to a \emph{short-ranged} squeezing matrix $\gamma^{\,AB}_{ij}=\lambda_{e(i,j)}\epsilon^{AB}$ whose only non-vanishing components are the edges $e(i,j)$ of the graph $\Gamma$. These states provide the first concrete realization of the conjecture on entanglement and the architecture of spacetime put forward in \cite{Bianchi:2012ev} as they show how a semiclassical geometry arises by gluing nearby quantum polyhedra with entanglement \cite{Baytas:2018wjd}. 

In order to distinguish between a microscopic area law and an effective area law, it is useful to introduce the notion of geometric entanglement entropy with a ``safety corridor'' around the region \cite{Casini:2008wt}: one considers a region $A$ containing $N_A$ nodes of the graph, an enlargement $AC$ that surrounds the region with a corridor with $N_C$ nodes, and the complement $B$. The quantity of interest is then the geometric mutual information (\ref{eq:mutual-information}) in the limit $N_A\to\infty$, $N_C\to\infty$, with $N_C/N_A\to 0$. In this limit, states that have only short-range correlations as Bell-network states satisfy a zero law for the mutual information. On the other hand, when the squeezing matrix has also \emph{long-ranged} non-vanishing components, for instance with an inverse-distance decay with respect to the average geometry \cite{Bianchi:2016tmw},  one finds an effective area law for the mutual information $S_{AB|C}(|\Gamma,\gamma\rangle)$ of squeezed vacua.

\section{Summary}
We presented recent developments at the interface of loop quantum gravity and quantum information, and discussed applications of entanglement measures to quantum geometry. In particular, after describing the Hilbert space of spin-network states together with its interpretation in terms of quantum geometries (Sec.~\ref{sec:spin-network}), we introduced the notions of link entanglement, intertwiner entanglement, and boundary spin entanglement (Sec.~\ref{sec:network-entanglement}). We then showed  how these notions encode the gluing of quanta of space and their relevance for the reconstruction of a quantum geometry from a network of entanglement structures (Sec.~\ref{sec:boundary-bulk}).
Moreover, using information theoretic bounds on the uncertainty of geometric observables and on their correlations (Sec.~\ref{sec:entropy-bounds}), we showed how the geometric entanglement entropy of spin-network states at fixed spins, treated as a many-body system of quantum polyhedra, allows us to identify a hierarchy of volume-law, area-law and zero-law states with different scaling laws for correlation functions (Sec.~\ref{sec:hierarchy}). In particular, we conjectured area-law states as the corner of the Hilbert space that encodes a semiclassical geometry (Sec.~\ref{sec:architecture}), and the geometric entanglement entropy as a probe of semiclassicality in quantum gravity.

The goal of these investigations is two-fold: (i) to determine the fundamental nature of spacetime geometry, by clarifying the role that entanglement plays in gluing spacetime quanta, and (ii) to provide new tools, both conceptual and numerical, for identifying the regime of loop quantum gravity where an effective description in terms of quantum fields on a classical spacetime is valid. These are a needed steps in order to extract robust observational predictions from loop quantum gravity.
While the results described in this chapter mostly focus on the kinematics,  we expect that these techniques can provide the basis for future investigations of the dynamics of loop quantum gravity. In the Hamiltonian framework, squeezed spin networks can provide a variational ansatz for the solution of the Hamiltonian constraint, with the variational parameters encoding directly the entanglement structure of the state. In the covariant framework, bulk reconstruction requires the entanglement structure in the boundary state to match the structure of the spinfoam dynamics.

\section*{Acknowledgments}
E.B.~acknowledges support from the National Science Foundation, Grant No.~PHY-2207851, and from the John Templeton Foundation via the ID 61466 grant, as part of the “Quantum Information Structure of Spacetime (QISS)” project (\hyperlink{http://www.qiss.fr}{qiss.fr})

%
%


\bibliographystyle{bib-style}
\bibliography{LQG-QI}

\end{document}